\documentclass[12pt,preprintnumbers,amsmath,amssymb]{article}

\usepackage{graphicx}
\usepackage{dcolumn}
\usepackage{bm}
\usepackage{subfigure,amsmath}
\newcommand{\be}{\begin{equation}}
\newcommand{\ee}{\end{equation}}
\newcommand{\ba}{\begin{eqnarray}}
\newcommand{\ea}{\end{eqnarray}}
\def\nonu{\nonumber}

\begin{document}
\begin{titlepage}

\begin{center}
\Large{Unparticle Physics and $A_{FB}^b$ on the $Z$ pole} \vskip 1cm

\normalsize {Mingxing Luo\footnote{Email: luo@zimp.zju.edu.cn},
             Wei Wu\footnote{Email: weiwu@zimp.zju.edu.cn} and
             Guohuai Zhu\footnote{Email: zhugh@zju.edu.cn} \\
\vskip .5cm
 {Zhejiang Institute of Modern Physics, Department of Physics, \\
 Zhejiang University, Hangzhou, Zhejiang 310027, P.R. China}
  \\
 }
\end{center}

\begin{abstract}
An attempt has been made to address the $3\sigma$ anomaly of the forward-backward asymmetry of $b$ quark in LEP data
via an unparticle sector.
For most part of the parameter space except certain particular regions,
the anomaly could not be explained away plausibly,
when constraints from other LEP observables are taken into account.
\end{abstract}

\end{titlepage}

\section{Introduction}

During the last 35 years or so, the Standard Model (SM) has been well tested by experiments.
In particular, the LEP experiment is one of the most impressive,
in which the statistical uncertainty is reduced by the huge number of events
and the systematic uncertainty reduced by the clean experiment environment.
The physical analysis was well documented. For example,
a recent report \cite{LEP} summarized all of the precision electroweak measurements on the $Z$ resonance.

Overall, the LEP data can be well interpreted by the SM, except for few small deviations.
One such conspicuous case is the forward-backward asymmetry of the bottom quark on the $Z$ resonance, $A_{FB}^{0,b}$,
which differs from the SM prediction by approximately three standard deviations.
Deviations as such may well be statistical fluctuations.
If it is not, there could be two remedies.
One possibility is high-order corrections within the SM,
which is, however, not supported by recent calculations \cite{radiativeSM}.
Another explanation is, of course, due to new physics effect \cite{newphy}.
In this paper, we will see to the possibility
whether the so-called unparticle sector would provide such an explanation.
Unfortunately, the answer turns out to be negative for most part of the parameter space.

The notion of unparticle sector was suggested recently by Georgi \cite{Ge1}.
It is supposed to be a hidden sector which has non-trivial conformal behavior at the low energy limit.
Its interaction with the SM sector is through an intermediate sector which is of high energy scale
and its effects may well appear at the TeV scale.
Non-trivial fixed point in the infrared are commonly used in condensed matter physics
to describe second order phase transitions, but rarely encountered in particle physics.
However, the existence of such infrared fixed points in Yang-Mills theories was realized many years ago \cite{bz}.
In a gauge theory with suitable number of massless fermions, one does have nontrivial infrared fixed points,
which ensures a non-trivial conformal sector in the infrared.
Actually, it was argued by Seiberg \cite{seiberg} that a conformal window could exist in supersymmetric
$SU(N_C)$ gauge theories with $N_f$ fermions if $3N_c/2<N_f<3N_c$.
But phenomenological implications of these exotic possibilities have not been addressed seriously.

Admittedly, one has yet to iron out a consistent framework for the unparticle physics
and there are many theoretical issues to be examined carefully.
For instance, it is well known that $S$-matrices cannot be defined in conformal field theories,
as one cannot define asymptotic states in these theories.
On the other hand, the unparticles must interact with the SM particles to be relevant,
but such interactions definitely break down the scale invariance.

Nevertheless, one may as well take such a novel framework as a working hypothesis
and then push forward to see how far it can take us.
In the framework of effective field theory,
one may discuss phenomenologies of the effective unparticle operators without
concrete understanding on their dynamics at high energy.
One interesting point is that \cite{Ge1},
thanks to the property of scale invariance, the unparticle looks like a non-integral number of
missing massless particles in the detector.
In this direction, there have been quite a bit activities to work out such implications
\cite{unpar-all}.

Interestingly, a queer phase \cite{Ge1} appears in the unparticle propagator in the time-like region,
which leads to novel interferences between unparticles and the SM processes
very different from the familiar pattern.\footnote{
This effect was also noticed in Drell-Yan process by Cheung {\it et al.} in \cite{unpar-all}.}
Potentially interesting phenomenologies could be observed at present or future colliders such as LEP and LHC.

In most cases, leading order effects of new physics are their interferences with the SM ones.
However, $Z$ pole is usually not a good place to see interference effects between the SM and new physics,
as amplitudes from the SM and new physics are out of phase by 90 degrees.
The extra phase factor in the unparticle propagator changes things drastically \cite{Ge1}.
The unparticle amplitude interferes with the SM one fully which may give a considerable contribution on the pole.
This provides a new opportunity to address the deviation of the forward-backward asymmetry of
$b$ quark in LEP measurements.
In this paper, we discuss this possibility systematically.

The paper is organized as follows:
In section \ref{SMdef}, several physical observables on the $Z$ resonance are discussed.
Section \ref{Unpar} introduces basic notions of an unparticle sector.
A detailed numerical analysis is presented in section \ref{Pheno}
on effects of the unparticles on physical observables measured in LEP,
with a particular emphasis on $A_{FB}^{0,b}$.
The results are finally summarized in section \ref{summary}.

\section{Physical Observables on the $Z$ Resonance \cite{langacker}}\label{SMdef}

In Standard Model, the differential cross section for $e^+e^-\rightarrow f \bar{f}$ through the $s$ channel is
  \ba
  \label{dcs}
  \frac{d\sigma}{d\cos\theta}&=&\frac{\beta s}{128\pi}[(|G_{LL}|^2+|G_{RR}|^2)(1+\beta
  \cos\theta)^2+(|G_{LR}|^2+|G_{RL}|^2)(1-\beta \cos\theta)^2 \nonu\\
  &+&2(1-\beta^2)\Re e(G_{LL} G_{LR}^*+G_{RR}G_{RL}^*)]~,
  \ea
and the total cross section is obtained by integrating out the $\theta$ angle.
The cross sections for scatterings of left- and right-handed electrons on unpolarized positrons are
  \be
  \sigma_X=\frac{\beta s}{64 \pi}[(|G_{XL}|^2+|G_{XR}|^2)(1+1/3 \beta^2)+2 (1-\beta^2)\Re e (G_{XL} G_{XR}^*)]~,
  \ee
where $s$ is the center of mass energy square, $\beta=(1-4 m^2/s)^{1/2}$, $m$ is the mass of the massive fermion,
and $G_{XY}$'s are
 \be
  \label{gxy}
  G_{XY}(s)= \displaystyle\sum_{A} g_X(A\rightarrow e^+e^-)g_Y(A\rightarrow f
\bar{f})*\Delta_A(s)~.
  \ee
Here $A$ is either $\gamma$ or $Z$; $X,\ Y =L$ or  $R$ are the chiralities of fermions, the propagators are
  \be
  \Delta_A(s)\{A=\gamma,Z\}=\left\{\frac{1}{s},\frac{1}{s-M_Z^2+iM_Z\Gamma_Z}\right\}~,
  \ee
and the couplings are
\ba
g_X(A\rightarrow f \bar{f})\{A=\gamma,Z\}&=&
\Bigr\{-eQ^f,\frac{-e}{\sin\theta_W \cos\theta_W}(I_3^f-Q^f\sin^2\theta_W)\Bigr\}~,
\ea
where $Q^f$ and $I_3^f$ are the electric charge and weak isospin of $f$, respectively.
Note that for left-handed (right-handed) fermions, $I_3^f$ are taken to be $\pm 1/2$ ($0$).

The hadronic cross section on the $Z$ resonance is
 \be
 \sigma_{had}^0=\displaystyle\sum_q\sigma(e^+ e^- \rightarrow q \bar{q})\simeq\sigma^0_b+2 \sigma^0_d+2 \sigma^0_u~,
 \ee
if $u,\ d\ ,s\ ,c$ quarks are regarded as massless.
Here and hereafter, the superscript $^0$ denotes quantities on the $Z$ resonance.

The left-right polarization asymmetry is defined as
  \be
  A_{LR}^0=\frac{\sigma^0_L-\sigma^0_R}{\sigma^0_L+\sigma^0_R}~,
  \ee
here the luminosity-weighted $e^-$ beam polarization magnitude is supposed to be 1.
The  $e^+ e^-$ final state is excluded here because it contains $t$-channel subprocess of photon exchange
which could dilute the result. $\mu^+ \mu^-$ and $\tau^+ \tau^-$ final states are considered in a complementary analysis \cite{LEP},
 which therefore will not be included in the following discussions.
With these selection rules, $A_{LR}^0$ is measured by summing hadronic final states in SLD experiment.
Thus,
  \be
  A_{LR}^0=\displaystyle\sum_q \left(\sigma_L^{0,q}-\sigma_R^{0,q}\right) / \sigma_{had}^0~.
  \ee

To distinguish relatively heavy flavors from light ones, one defines
  \be
  R_b^0=\frac{\sigma^0_b}{\sigma_{had}^0}, \ \ \ \ \ \
  R_c^0=\frac{\sigma^0_c}{\sigma_{had}^0}~.
  \ee
Finally, one defines the forward-backward asymmetry of $f\bar{f}$ production on the $Z$ resonance:
  \be
  A_{FB}^{0,f}=\frac{\sigma^{0,f}_F-\sigma^{0,f}_B}{\sigma^{0,f}_F+\sigma^{0,f}_B}~,
  \ee
  where
  \be
  \sigma^{0,f}_F=\int_{0}^{\pi/2}\frac{d\sigma^0_f}{d\cos\theta} d\cos\theta, \ \ \ \ \ \
  \sigma^{0,f}_B=\int_{\pi/2}^{\pi}\frac{d\sigma^0_f}{d\cos\theta} d\cos\theta~.
  \ee

Shown in table \ref{Tabel:Measurements} are the latest experiment data
and SM global fit \cite{LEP} of these physical observables. All observables are well consistent with the
SM except $A_{FB}^{0,b}$, which deviates by almost three standard deviations.

\begin{table}
\begin{center}
  \begin{tabular}{@{ } c c c@{ }}
  \hline
  \hline
  \\
  \ &Measurement&\ SM fit\\
  \hline
  \\
  $\sigma_{had}^0$(nb)\ \ &$41.540\pm 0.037$\ \ &$41.481\pm 0.014$\\
  \\
  $R_b^0$\ \ &$0.21629\pm 0.00066$\ \ &$0.21562\pm 0.00013$\\
  \\
  $R_c^0$\ \ &$0.1721\pm 0.0030$\ \ &$0.1723\pm 0.0001$\\
  \\
  $A_{LR}^0$(SLD)\ \ &$0.1514\pm 0.0022$\ \ &$0.1480\pm 0.0011$\\
    \\
  $A_{FB}^{0,l}$\ \ &$0.0171\pm 0.0010$\ \ &$0.01642\pm 0.00024$\\
  \\
  $A_{FB}^{0,c}$\ \ &$0.0707\pm 0.0035$\ \ &$0.0742\pm 0.0006$\\
  \\
  $A_{FB}^{0,b}$\ \ &$0.0992\pm 0.0016$\ \ &$0.1037\pm 0.0008$\\
  \hline
  \hline
  \end{tabular}
  \end{center}
 \caption{\label{Tabel:Measurements}
Physical observables:
measurements on the $Z$ resonance in the second column and SM global fits in the third column \cite{LEP}.}
\end{table}

\section{The Unparticle Sector}\label{Unpar}

Interactions of vector-like unparticle with SM fermions can be approximated by an effective Lagrangian
  \be
  \mathcal{L}_{int}=\frac{c^f_{V\mathcal{U}}}{M_Z^{(d_{\mathcal{U}}-1)}}\bar{f}\gamma_{\mu}f
  \mathcal{U}^{\mu}_V+\frac{c^f_{A\mathcal{U}}}{M_Z^{(d_{\mathcal{U}}-1)}}\bar{f}\gamma_{\mu}\gamma_5 f
  \mathcal{U}^{\mu}_A~.
  \ee
Following conventions in ref. \cite{Ge1},
the couplings $c^f_{V\mathcal{U}}$ and $c^f_{A\mathcal{U}}$ are normalized in terms of the the $Z$ boson mass.
Scalar unparticles may also couple to the SM fermions and thus affect $A_{FB}^{0,b}$.
The derivation are actually very similar to the case of vector unparticles,
though numerically they might be different.
However as discussed recently by Fox {\it et al.} in \cite{unpar-all}, the scale invariance of the unparticles may
break down if scalar unparticles are coupled to the Higgs. Even if such coupling does not exist at tree level,
it could be regenerated through loop diagrams. Therefore we choose not to discuss scalar unparticles in the following.

One hopes that the unparticle sector could account for the roughly $3\sigma$ deviation between the SM prediction
and the LEP measurement on $A_{FB}^{0,b}$.
On the other hand, unparticles should not affect other observables too much
so as not to invalidate agreements between the SM global fit results and their LEP measurements.
Thus, $c^f_{V\mathcal{U}}$ and $c^f_{A\mathcal{U}}$ have to be flavor-dependent.
For simplicity, we assume that the unparticle couplings with the SM fermions are universal
except those with the $b$ quark,
\begin{equation}\label{couplings}
c^f_{V\mathcal{U}}=\left \{
\begin{aligned}
& ~~c_{V\mathcal{U}}~~~(f \neq b)~, \\ & \lambda c_{V\mathcal{U}} ~~~(f = b)~;
\end{aligned}
\right .
\hspace*{1.5cm}
c^f_{A\mathcal{U}}=\left \{
\begin{aligned}
& ~~c_{A\mathcal{U}}~~~(f \neq b)~, \\ & \lambda c_{A\mathcal{U}} ~~~(f = b)~;
\end{aligned}
\right .
\end{equation}
$|\lambda| > 1$ if one wishes to address the $A_{FB}^{0,b}$ deviation.

Following \cite{Ge1}, the vector-like unparticle operators are assumed to be transverse
and the propagator is given by
  \ba
  \label{prop}
  \int d^4 x
  e^{iPx}<0|T\mathcal{U}^{\mu}_{V(A)}(x)\mathcal{U}^{\mu}_{V(A)}(0)|0>
  =i\frac{A_{d_{\mathcal{U}}}}{2}\frac{-g^{\mu\nu}+ P^{\mu}P^{\nu}/P^2}{\sin(d_{\mathcal{U}}\pi)}
  \Bigr(-P^2-i\epsilon\Bigr)^{d_{\mathcal{U}}-2}~,
  \ea
with
  \be
  A_{d_{\mathcal{U}}}=\frac{16\pi^{5/2}}{(2 \pi)^{2
  d_{\mathcal{U}}}}\frac{\Gamma(d_{\mathcal{U}}+1/2)}{\Gamma(d_{\mathcal{U}}-1)\Gamma(2 d_{\mathcal{U}})}~.
  \ee

It is then straightforward to calculate unparticle contributions to the process $e^+e^-\rightarrow f \bar{f}$.
Following procedures in section \ref{SMdef} and define
\begin{equation}
\Delta_{\mathcal{U}}(s)=\frac{A_{d_{\mathcal{U}}}}{2}
\frac{\left ( -P^2-i\epsilon \right )^{d_{\mathcal{U}}-2} }{\sin(d_{\mathcal{U}}\pi)}~,\hspace*{1cm}
g_{L,R}(\mathcal{U}\to f \bar{f})=\left \{
\begin{aligned}
& ~~\frac{c_{V\mathcal{U}} \mp c_{A\mathcal{U}}}{M_Z^{d_{\mathcal{U}}-1}}~~~(f \neq b)~,
\\ & \lambda \frac{c_{V\mathcal{U}} \mp c_{A\mathcal{U}}}{M_Z^{d_{\mathcal{U}}-1}} ~~~(f = b)~.
\end{aligned}
\right .
\end{equation}
The unparticle contributions are taken into account
by letting $A=\gamma$, $Z$, and $\mathcal{U}$ in Eqs. (\ref{dcs}-\ref{gxy}).

At Z pole, the SM amplitude is almost pure imaginary while normally the new physics contribution is real. Therefore
it is hard to observe interference effects at or near Z pole. However
as first discussed by Georgi in the second paper of \cite{Ge1},
the phase $e^{-i(d_{\cal U}-2)\pi}$ in Eq. (\ref{prop}) causes the unparticle amplitude to be complex
in the time-like region and thus
provides a novel possibility for unparticles to interfere with the SM amplitude at the $Z$ resonance.
Note that the unparticle sector introduces four free parameters: $c_{A\mathcal{U}}$,$c_{V\mathcal{U}}$, $\lambda$,
and $d_{\mathcal{U}}$.

\section{Phenomenological Analysis}\label{Pheno}

We now discuss unparticle contributions to physical observables on the $Z$ resonance and compare them with the LEP data.
Since our main concern is about $A_{FB}^{0,b}$,
we shall first consider the influence of the unparticle sector on this quantity.

\begin{figure}[tb]
    \subfigure[]{
    \label{fig:afb4case:subfig:a}
    \begin{minipage}[b]{.47\textwidth}
    \centering
    \includegraphics[scale=0.7]{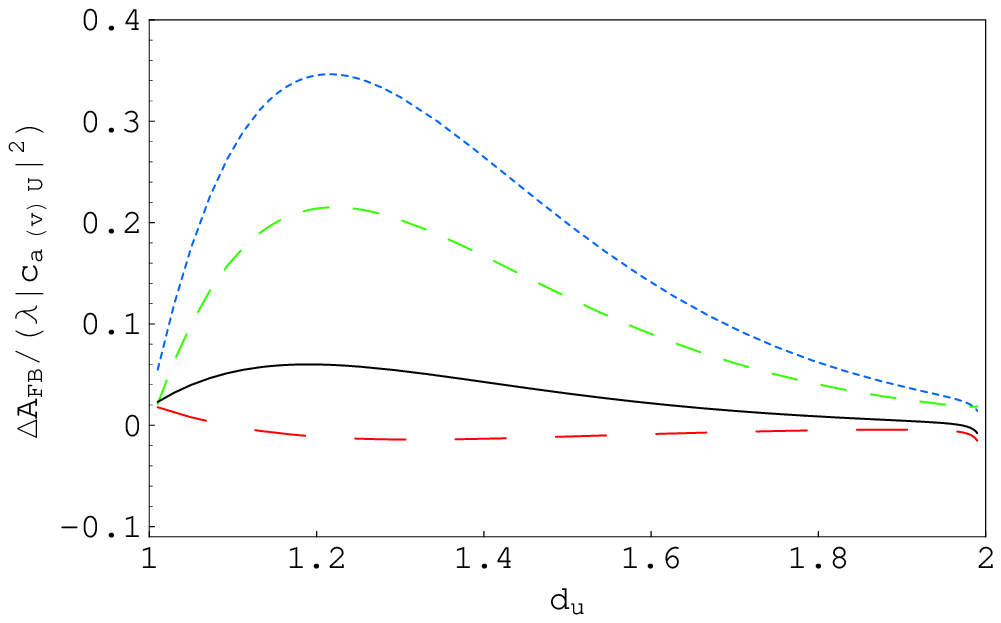}
    \end{minipage}}
    \subfigure[]{
    \label{fig:afb4case:subfig:b}
    \begin{minipage}[b]{.47\textwidth}
    \centering
    \includegraphics[scale=0.7]{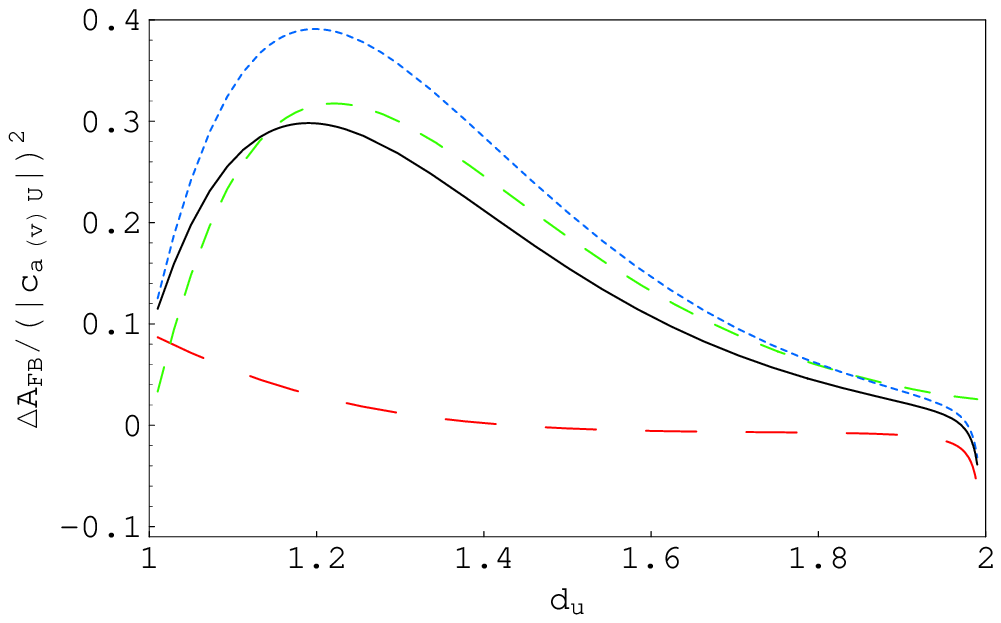}
    \end{minipage}}
\caption{\label{fig:afb4case}
Changes of forward-backward asymmetries for (a) $e^+ e^- \rightarrow b \bar{b}$ in unit of
$\lambda \vert c_{A(V){\cal U}} \vert^2$ and
 (b) $e^+ e^- \rightarrow \ell \bar{\ell}$ in unit of $\vert c_{A(V){\cal U}} \vert^2$
 versus $d_{\mathcal{U}}$ on the $Z$ pole.
 The solid, dot, short-dashed and long-dashed lines represent cases of (i) $c_{A\mathcal{U}}=c_{V\mathcal{U}}$,
 (ii) $c_{A\mathcal{U}}=-c_{V\mathcal{U}}$, (iii) $c_{V\mathcal{U}}\neq 0, c_{A\mathcal{U}}=0$ and
 (iv) $c_{A\mathcal{U}}\neq 0, c_{V\mathcal{U}}=0$, respectively.}
\end{figure}

For convenience, one may express the forward-backward asymmetry in the massless limit
in terms of vector and axial-vector couplings\footnote{The formula in massless limit is quoted solely for the purpose of the qualitative discussions.
 The b quark mass effects are taken into account in the numerical analysis. }
  \be
  \label{VA}
  A_{FB}=\frac{3}{2}\Bigr(\frac{\Re e(G_{VV}^*G_{AA})+\Re e(G_{VA}^*G_{AV})}{|G_{VV}|^2+|G_{AA}|^2+|G_{VA}|^2+|G_{AV}|^2} \Bigr)~.
  \ee
The definition of $G_{xy}$ with $x,y=V$ or $A$ is the same as those for the left and right ones in Eq. (\ref{gxy}),
with $g_{V,A}=(g_R \pm g_L)/2$.

Shown in Figure \ref{fig:afb4case} are changes of forward-backward asymmetries of the $b$ quark
and leptons on the $Z$ pole, due to the unparticle sector. For example,
\[
       A_{FB}^\ell=A_{FB}^{\ell, SM}+ x*\vert c_{A(V){\cal U}} \vert^2 + {\cal O}(\vert c_{A(V){\cal U}} \vert^4)~,
\]
and the coefficient $x$ denotes the change of $A_{FB}^\ell$ in unit of $\vert c_{A(V){\cal U}} \vert^2$,
as shown in Figure \ref{fig:afb4case}, while the ${\cal O}(\vert c_{A(V){\cal U}} \vert^4)$ term is neglected.

At the resonance, the QED amplitude is very small compared with the weak one.
This leads to the ordering pattern of the SM amplitude: $G_{AA} \gg G_{VA}, G_{AV} \gg G_{VV}$.
Thus, the leading interference effect between the unparticle sector and the SM amplitude arises from
the term $\Re e(G_{AA}^{SM} G_{VV}^{\mathcal{U}~*})$.
Consequently, the contribution from the unparticle coupling $c_{V{\mathcal{U}}}$ is the largest.
In Fig.\ref{fig:afb4case},
the long-dashed line with $c_{V\mathcal{U}}=0$ has almost negligible effect on $A_{FB}$.
Note also that the unparticle contribution to $A_{FB}^{0,\ell}$ is $\lambda$-independent,
as seen from Fig. \ref{fig:afb4case}b.
To be consistent with experimental observations,
one hopes the change of $A_{FB}^{0,b}$ to be relatively large to interpret the deviation.
On the other hand, the change of $A_{FB}^{0,\ell}$ should be small enough, say within the $1\sigma$ experimental error.
Similarly, constraints on the unparticle couplings could also be obtained from the forward-backward asymmetry
of the $c$ quark,
which however are less restricted compared with those from the leptons and therefore not shown here.

\begin{figure}[tb]
\centerline{\includegraphics[width=15cm]{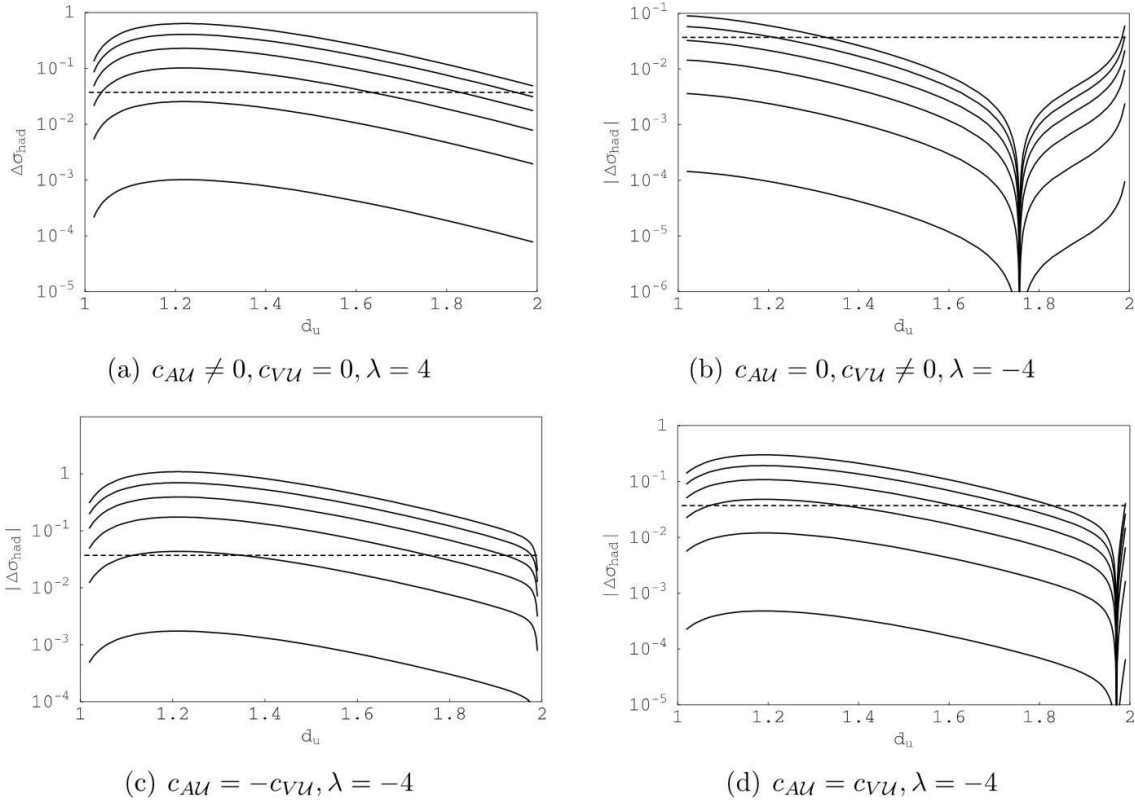}}
\caption{\label{fig:cs}
Contributions to hadronic cross section on the $Z$ pole from the unparticle sector versus $d_{\mathcal{U}}$.
For illustration, $|\lambda|$ is chosen to be 4.
The minus sign of $\lambda$ in cases (b), (c), (d) is chosen so as to have negative contributions to $A_{FB}^{0,b}$,
as required by the LEP data.
In case (a), for any given sign of $\lambda$, the contribution to $A_{FB}^{0,b}$ oscillates in sign with $d_{\mathcal{U}}$.
The positive sign as chosen is for illustration, but our conclusion does not depend on this particular choice.
Solid lines from top down show different inputs for $c_{A(V)\mathcal{U}}=0.25$, $0.2$, $0.15$, $0.1$,
$0.05$, and $0.01$, respectively.
Horizental dashed lines denote limits of $1\sigma$ experimental errors.}
\end{figure}

The unparticle sector also affects other physical observables.
We now discuss its impact on the hadronic cross section $\sigma_{had}^0$,
the ratios $R_b^0$ and $R_c^0$, and the left-right polarization asymmetry $A_{LR}^0$.
As seen from Table \ref{Tabel:Measurements}, measurements on these quantities are well consistent with the SM fits.
Constraints on the unparticle couplings can be obtained by these observables.

For the hadronic cross section,
\[
 \sigma_{had}^0 \propto |G_{VV}|^2+|G_{AA}|^2+|G_{VA}|^2+|G_{AV}|^2~.
\]
Again, the $|G_{AA}|^2$ term dominants on the $Z$ resonance in the SM.
The leading interference term
between the unparticles and the SM part is proportional to $\Re e(G_{AA}^{SM} G_{AA}^{\mathcal{U}~*})$.
Therefore, $\sigma_{had}^0$ is quite sensitive to $c_{A\mathcal{U}}$, but much less so to $c_{V\mathcal{U}}$,
as shown in Fig.\ref{fig:cs}, where the ratio $\Delta \sigma_{had}=(\sigma_{had}-\sigma_{had}^{SM})/\sigma_{had}$
represents the difference for the hadronic cross section with or without unparticle contributions.
Here and hereafter, we choose four typical scenarios to be examined,
in which the unparticle couplings are vector, axial-vector, left-handed and right-handed, respectively.
At the quark level, $\sigma_{had}^0$ contains the production of $u$, $d$, $s$, $c$ and $b$ quarks
but only the $b$ quark part depends on $\lambda$.
Therefore the constraints obtained from $\sigma_{had}^0$ is not sensitive to the value of $\lambda$.
Note also that in Fig.\ref{fig:cs}, $\vert \Delta \sigma_{had} \vert$ is plotted on a logarithmic scale.
The dips in Fig. (2b), (2d) means that the unparticle contributions vanish at these specific
$d_{\cal U}$ values, and $\Delta \sigma_{had}$ changes sign across these dips..

\begin{figure}[bt]
\centerline{\includegraphics[width=15cm]{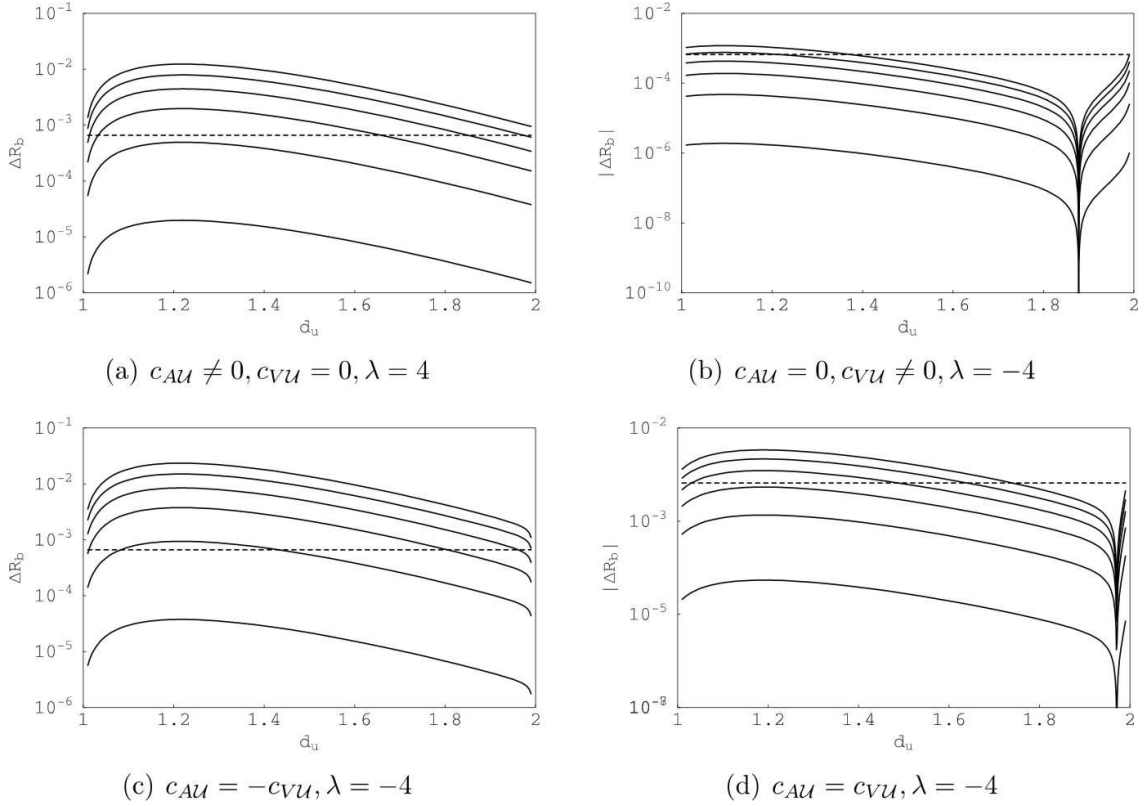}}
\caption{\label{fig:Rb}
Unparticle contributions to $R_b$ on the $Z$ pole versus $d_{\mathcal{U}}$.
Parameters and conventions are the same as those in Fig. \ref{fig:cs}. }
\end{figure}

$R_b^0$ and $R_c^0$ are both measured precisely at LEP experiments, as shown in Table \ref{Tabel:Measurements}.
The experimental error of $R_b^0$ is about $5$ times smaller than that of $R_c^0$.
Naturally, one anticipates that $R_b^0$ gives stricter constraint on the unparticle couplings than $R_c^0$,
which is verified by our numerical analysis.
Thus, only results from $R_b^0$ are plotted in Fig. \ref{fig:Rb}.
By similar reasoning as those for $\sigma_{had}^0$,
$R_b^0$ is quite sensitive to $c_{A\mathcal{U}}$ but much less so to $c_{V\mathcal{U}}$.
Note that $\lambda$ appears only in the unparticle coupling with $b$ quarks,
so the $\lambda$ dependence appears in the change of $R_b$ only.

\begin{figure}[tb]
\centerline{\includegraphics[width=15cm]{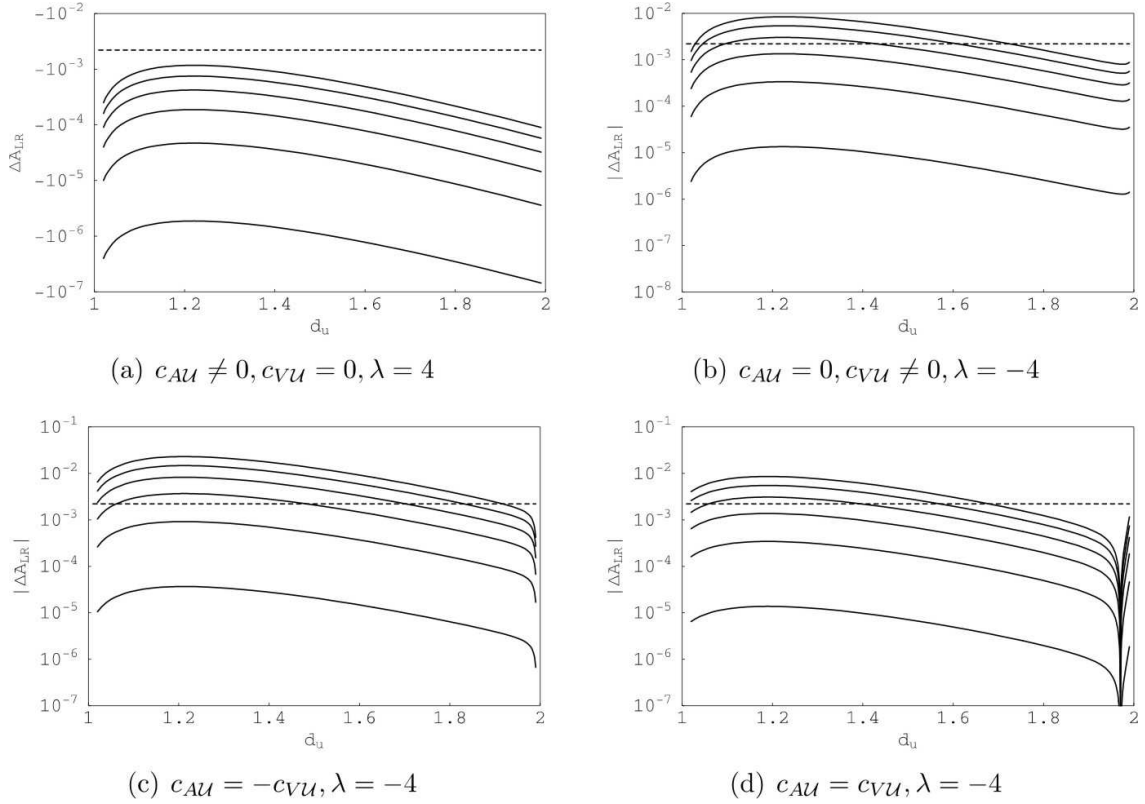}}
\caption{\label{fig:eeLR}
Unparticle contributions to Left-Right asymmetry on the $Z$ pole versus $d_{\mathcal{U}}$.
Parameters and conventions are the same as those in Fig. \ref{fig:cs}. }
\end{figure}

Finally we consider the left-right polarization asymmetry $A_{LR}^{0}$.
Contrary to the cross section-related observables, such as $\sigma_{had}^0$ and $R_b$,
$A_{LR}^{0}$ could lead strong constraints on the parameter $c_{V\mathcal{U}}$ instead of $c_{A\mathcal{U}}$, as shown
in Fig.\ref{fig:eeLR}.
Since the quoted value of $A_{LR}^{0}$ is measured by SLD experiment by summing over all hadronic final states,
just like the case of $\sigma_{had}^0$, the curves in Fig.\ref{fig:eeLR} are not sensitive to the value of $\lambda$.

\begin{figure}[tb]
\centerline{\includegraphics[width=15cm]{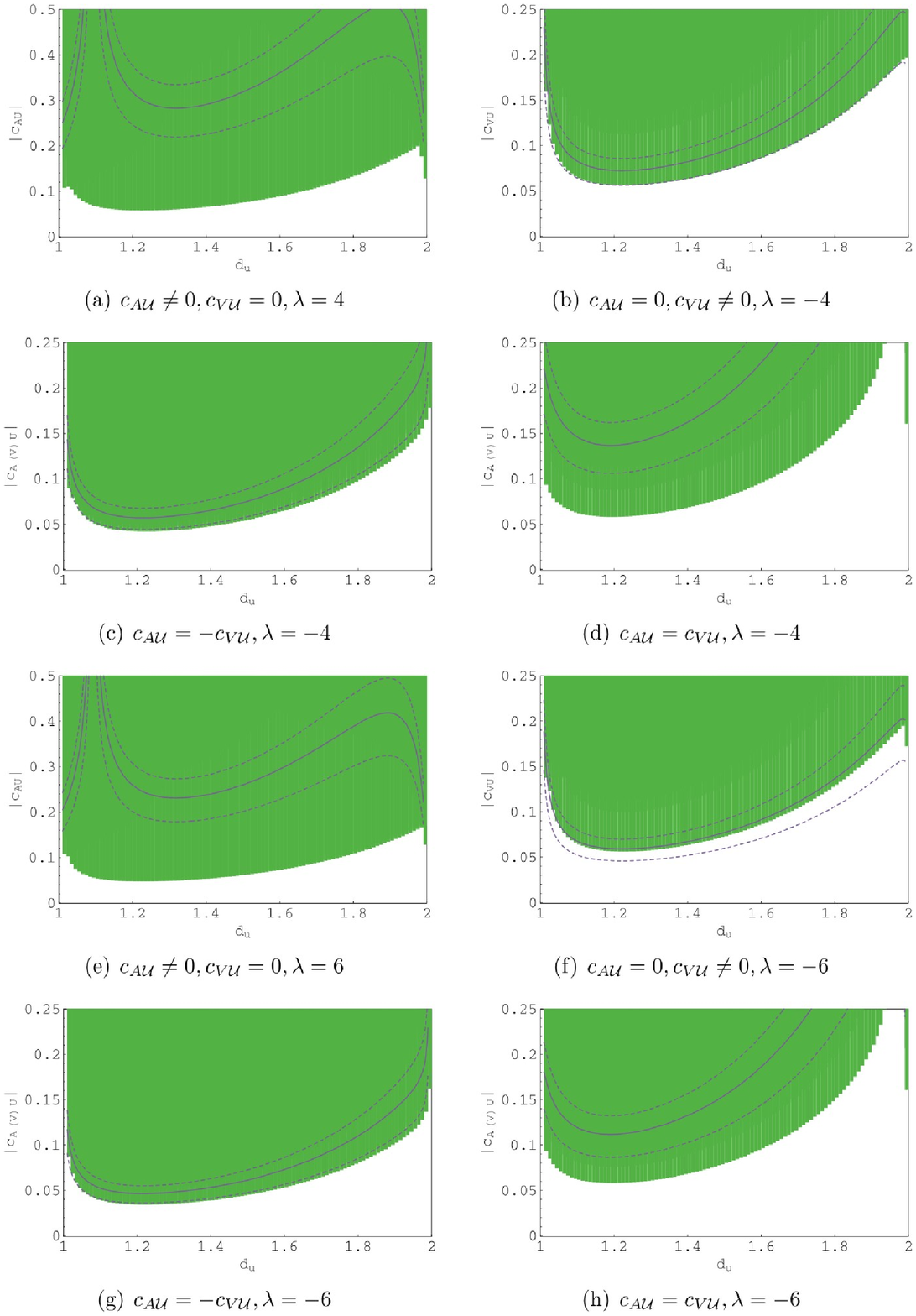}}
\caption{\label{fig:para}
Scanning the parameter space of the unparticle couplings by including all the physical observables
discussed in the paper.  Here $|\lambda|=4$, $6$ are taken for illustrations. The solid lines represent that
the unparticle sector could provide the central-value difference between the LEP measurements and the SM fit of
$A_{FB}^{0,b}$. The dashed lines take into account the $1\sigma$ experimental error.
The grey (green) areas are excluded by the combined analysis from other observables which are well consistent with
the SM fits.}
\end{figure}

Now we are ready to combine all constraints obtained from $A_{FB}^{0,\ell}$, $\sigma_{had}^0$, $R_b^0$
and $A_{LR}^{0}$,
to see whether there are areas in the parameter space to interpret the observed deviation of $A_{FB}^{0,b}$.
Again four typical scenarios under consideration are chosen to be axial-vector, vector, left-handed and right-handed
couplings between the unparticles and the SM part.
Corresponding results are drawn in Fig. \ref{fig:para}(a)-(d), respectively.
It seems difficult for the unparticle sector to be able to account for the observed deviation on $A_{FB}^{0,b}$.
Note that constraints from $R_b^0$ have similar $\lambda$-dependence as those from $A_{FB}^{0,b}$.
Although only the cases with $\lambda=4$ and $6$ are shown in Fig. \ref{fig:para},
it is not difficult to go through detailed numerical investigations
to check that the scenarios with axial-vector, left-handed and right handed couplings
(Fig. \ref{fig:para}(a),(c),(d),(e),(g),(h)) are
almost completely excluded for any reasonable value of $\lambda$.
However, the scenario with pure vector coupling (Fig. \ref{fig:para}(b),(f)) is more subtle:
Here the most stringent constraint is from $A_{FB}^{0,\ell}$, which is independent on $\lambda$.
Therefore for a larger $\lambda$,
the curves of $A_{FB}^{0,b}$ may be lowered and the anomaly in $A_{FB}^{0,b}$ can be explained away.
For example, if $\lambda \ge 6$, a large pure vector unparticle coupling with $b$ quarks
$c_{V\mathcal{U}}^b > 0.3$ might reduce the theoretical prediction on $A_{FB}^{0,b}$
to be consistent with the LEP data (Fig.\ref{fig:para}(f)).
In general, in the particular region where the unparticle coupling with $b$ quarks is predominately vector-like
and $\lambda$ is substantially larger than 1,
the anomaly in $A_{FB}^{0,b}$ seems to be explained away.
Whether this provides a plausible solution indeed, it heavily relies on one's taste.
Some may argue that it is a little far stretched.

Off-resonance data, as those discussed by Bander {\it et al.} and Cheung {\it et al.} in \cite{unpar-all},
provides similar but weaker constraints in most cases,
which are not included here as they are not particularly illuminating.

\section{Summary}\label{summary}

In this paper, we have discussed the possibility whether the $3\sigma$ deviation of the forward-backward asymmetry of
$b$ quark between the LEP measurements and the SM fit could be accounted for by an unparticle sector.
By considering constraints from other observables, namely the hadronic cross section $\sigma_{had}^0$,
the ratio $R_b$, the left-right asymmetry $A_{LR}^0$ and the leptonic forward-backward asymmetry $A_{FB}^{0,\ell}$,
which are all well consistent with the SM fits,
it seems quite difficult to explain the $A_{FB}^{0,b}$ anomaly by the notion of unparticles.
Specifically, if the unparticle couplings with the SM fermions are axial-vector,
left-handed or right-handed, it is almost impossible to interpret the $3\sigma$ deviation of $A_{FB}^{0,b}$.
In the particular region where the unparticle coupling with $b$ quarks is predominately vector-like
and $\lambda$ is substantially larger than 1,
the anomaly in $A_{FB}^{0,b}$ seems to be explained away.
Whether this provides a plausible solution indeed, it heavily relies on one's taste.

\section*{Acknowledgments}
 This work is supported in part by the National Science Foundation of China under grant No.10425525 and No.10645001.

\end{document}